\documentclass[conference]{IEEEtran}
\usepackage{cite}
\usepackage{amsmath,amssymb,amsfonts}
\usepackage{gensymb}

\usepackage{algorithmic}
\usepackage{graphicx}
\usepackage{textcomp}
\usepackage{xcolor}

\def\BibTeX{{\rm B\kern-.05em{\sc i\kern-.025em b}\kern-.08em
    T\kern-.1667em\lower.7ex\hbox{E}\kern-.125emX}}

\begin{document}

\title{Frequency Diverse Array OFDM System for Joint Communication and Sensing}

\author{
    \IEEEauthorblockN{Marcin Wachowiak\IEEEauthorrefmark{1}\IEEEauthorrefmark{2}, 
    André Bourdoux\IEEEauthorrefmark{1}, 
    Sofie Pollin\IEEEauthorrefmark{2}\IEEEauthorrefmark{1}}
    \IEEEauthorblockN{
        \IEEEauthorrefmark{1} imec, Kapeldreef 75, 
        3001 Leuven, Belgium \\
        \IEEEauthorrefmark{2} Department of Electrical Engineering, 
        KU Leuven, Belgium \\
        Email: marcin.wachowiak@imec.be 
        }   
}

\maketitle
 
\begin{abstract}
The frequency-diverse array (FDA) offers a time-varying beamforming capability without the use of phase shifters. The autoscanning property is achieved by applying a frequency offset between the antennas. This paper analyzes the performance of an FDA joint communication and sensing system with the orthogonal frequency-division multiplexing (OFDM) modulation. The performance of the system is evaluated against the scanning frequency, number of antennas and number of subcarriers. The utilized metrics; integrated sidelobe level (ISL) and error vector magnitude (EVM) allow for straightforward comparison with a standard single-input single-output (SISO) OFDM system.

\end{abstract} 

\begin{IEEEkeywords}
Frequency diverse array (FDA), frequency-modulated array (FMA), orthogonal frequency-division multiplexing (OFDM), integrated sensing and communications (ISAC)
\end{IEEEkeywords}

\section{Introduction}

\subsection{Problem Statement}
The move towards millimeter-wave communications and sensing requires an extensive utilization of antenna arrays to compensate for the free space path loss \cite{mmw_comms_it_will_work}. Ideally, the number of signal chains has to scale according to the number of antennas. However, that would translate into a linear increase in the cost and power consumption of the system. The constant effort to reduce the cost of wireless systems, to enable their massive deployment, has led to the concept of frequency-diverse arrays (FDA) (also known as frequency-modulated arrays (FMA)). The FDA architecture is similar to that of the analog phased array (PA). In a phased array, each antenna is preceded by an analog phase shifter and is connected to a single shared RF chain via a splitter (TX) or combiner (RX). In the FDA architecture, a phase shifter is replaced by a mixer, which introduces a frequency offset between signals transmitted by each antenna. The FDA concept offers yet another way to manipulate and explore the spatial dimension of the radiated signals.

\begin{figure}[t]
    \centering
    \includegraphics[width=\columnwidth]{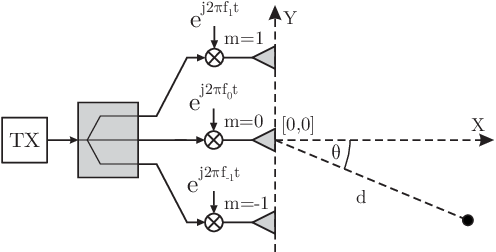}
    \caption{Considered frequency diverse array system. Each antenna is fed with the same signal shifted in frequency. The reference of the coordinate system is set to the center of the array. $d$ is the distance to a point of interest and $\theta$ is the azimuth angle in regard to the array normal.}
    \label{fig:fda_diagram}
\end{figure}

\subsection{Relevant works}
The frequency-diverse arrays have been mostly investigated in the radar context \cite{fda_tma_overview}, \cite{fda_mimo_techniques},\cite{fda_beampattern_analysis}. However, the analyses were limited to pulsed or narrowband cases and did not consider interference between the transmitted signals due to the overlap in the frequency domain. 
In \cite{fda_lfm_experimentation} the concept of FDA radar with linear frequency modulated (LFM) continuous-wave (CW) signals was validated experimentally, showing the feasibility of the system. The work did not address the communication aspect of the FDA and experienced a loss in effective sensing bandwidth due to the utilization of the FDA scheme with an LFM waveform. 
The OFDM FDA communication scheme where the subcarriers of the OFDM symbol were split between antennas was analyzed in \cite{fda_ofdm_sc_per_antenna}. The scheme was investigated in the context of secure communications and requires a separate signal chain per antenna.
A joint communications and sensing system with orthogonal chirps was presented in \cite{fda_ofdm_jcns_chirped}. It also requires a separate signal chain per antenna and offers a low communication rate.
In \cite{fma_tx_jcns}, a switched FDA and PA operation was investigated. The FDA mode was used to detect the target/receiver, whose location was later used during communications mode with a phased array.
A joint communications and sensing system based on frequency index modulation with an FDA was considered in \cite{fda_jcns_index_modulation}. The discussed work employed narrowband signals and required complex processing at the receiver.

\subsection{Contributions}
In this paper, the FDA is combined with wideband OFDM, where each antenna transmits all the subcarriers of the OFDM symbol. The frequency offset between the antennas is much smaller than the signal bandwidth, which results in interference between the symbols, both for the communication and radar modes. The time-varying beam pattern is investigated and its effects on the sensing and communication performances are discussed. Selected performance metrics, which are integrated sidelobe level (ISL) and error vector magnitude (EVM), allow for a straightforward comparison between the standard single-input single-output (SISO) OFDM communication and sensing system.

\section{System model}

\subsection{Frequency Diverse Array}
Consider a uniform linear array (ULA) with $M$ elements as shown in Fig. \ref{fig:fda_diagram}. The indexing of the antennas is $m\in \{0, \pm1, \ldots, \pm(M-1)/2\}$ for odd $M$ and  $m\in \{\pm0.5, \pm1.5, \ldots, \pm(M-1)/2\}$ for even $M$. Each antenna transmits the same baseband signal at a distinct frequency given by 
\begin{equation}
    f_m = f_c + m\Delta f,
    \label{eq:ant_freq_offset}
\end{equation}
where $f_c$ is the center operating frequency and $\Delta f$ is the frequency offset between the antenna elements. 
The array is assumed to operate in the far field with the center of the array selected as the reference point. The path difference in the direction $\theta$ between the $m$th antenna and the reference is then
\begin{equation}
    \Delta d_m = m d_{\mathrm{a}} \sin{(\theta)},
\end{equation}
where $d_{\mathrm{a}}$ denotes the array inter-element spacing.
Due to the symmetric indexing of $m$, the operating frequencies of antennas $f_m$ are symmetrically and uniformly distributed around the $f_c$. The same holds for the path differences $\Delta d_m$, which are centered around $0$. 

\begin{figure}[t]
    \centering
    \includegraphics[width=\columnwidth]{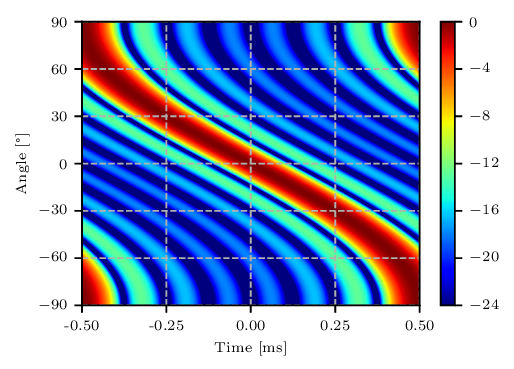}
    \caption{Evolution of the FDA beam pattern and normalized gain [dB] over time for ULA with $M=8$ antennas  and $\Delta f=1$ kHz.}
    \label{fig:fda_time_angle_beampattern}
\end{figure}

Because of the different operating frequencies, the shared path distance across antennas to some point of interest in space does not introduce a common phase shift. Therefore the total distance to the point of interest has to be taken into account. Considering a point of interest at a distance $d$ and angle $\theta$ the total phase shift between the $m$th antenna and the reference is
\begin{align}
    \label{eq:fda_mth_phshift}
    \Delta \phi_{m} =& 2\pi \left( \frac{\left( f_c + m\Delta f \right) \left(d + \Delta d_m \right) }{c}  - \frac{f_c d}{c} \right) \\
    =& \frac{2\pi}{c} \Bigl( m \Delta f d + m^2 \Delta f d_{\mathrm{a}} \sin{(\theta )}
     + f_c m d_{\mathrm{a}} \sin{(\theta)} \Bigr), \nonumber 
\end{align}
where $c$ is the speed of light. 
To obtain a closed-form expression of the array factor (AF) the quadratic term in \eqref{eq:fda_mth_phshift} has to be either approximated \cite{fda_tma_overview} or neglected in the derivations \cite{fda_mimo_techniques}. Given the maximum frequency offset $\Delta f (M - 1)/2 \ll f_c$ is much smaller than the carrier frequency, the contribution of the quadratic term is considered negligible. Substituting the range dependency with $d = t c$ the total phase shift between the antenna elements is
\begin{equation}
    \label{eq:fda_mth_phshift_simplified}
    \Delta \phi_m \approx 2\pi m \left( \Delta f t + \frac{d_{\mathrm{a}}}{\lambda_{\mathrm{c}}}  \sin{(\theta)} \right),
\end{equation}
where $\lambda_{\mathrm{c}} = c / f_c$. Given that the power is uniformly split across antennas the  array factor of the FDA is given by
\begin{align}
   \label{eq:fda_af}
   \mathrm{AF_{FDA}}(\theta, t) &= \frac{1}{\sqrt{M}} \sum_{m=-(M-1)/2}^{(M-1)/2}{e^{j 2\pi m \left(\Delta f t + \frac{d_{\mathrm{a}}}{\lambda_{\mathrm{c}}}  \sin{(\theta)} \right)}} \nonumber \\
   &= \frac{1}{\sqrt{M}} \frac{\sin{ \left( M \pi \left( \Delta f t + \frac{d_{\mathrm{a}}}{\lambda_{\mathrm{c}}}\sin{(\theta)} \right) \right) } }
   {\sin{ \left( \pi \left( \Delta f t + \frac{d_{\mathrm{a}}}{\lambda_{\mathrm{c}}}\sin{(\theta)} \right) \right) }}.
\end{align}
In contrast to conventional PAs, the time-varying phase shift due to different operating frequencies of the antennas introduces a time-varying beam pattern. The beamforming angle of FDA at time $t$ is
\begin{equation}
    \label{eq:fda_angle_vs_time}
    \theta(t) = -\sin^{-1}{\left( \frac{\lambda_{\mathrm{c}}}{d_{\mathrm{a}}} \Delta f t \right)}.
\end{equation}
The periodicity of the beam pattern in time is $1 / \Delta f$ with lobe width $2 / (M \Delta f)$. The lobe width and periodicity of the beam pattern in angle are the same as for a standard ULA \cite{phased_array_handbook}. An example of an FDA beam pattern evolution over time is shown in Fig. \ref{fig:fda_time_angle_beampattern} for an 8-element ULA with the frequency offset of $1$ kHz.

\vspace{-15pt}
\section{System analysis}
To investigate the limitations imposed by the FDA autoscanning property, a system analysis is performed. Considering an OFDM system with $K$ subcarriers and bandwidth $B$, the subcarrier spacing is $\Delta f_{\mathrm{sc}} = B/K$. Given these parameters, the duration of an OFDM symbol is 
\begin{equation}
    \label{eq:ofdm_symb_dur}
    T_{\mathrm{s}} = \frac{K + N_{\mathrm{cp}}}{B},
\end{equation}
where $N_{\mathrm{cp}}$ is the number of samples of the cyclic prefix (CP). The cyclic prefix is taken into account as the system is also considered for communications.
The ULA angular resolution can be approximated by \cite{balanis_antenna_book}
\begin{equation}
    \label{eq:angular_res}
    \Delta \Theta \approx \frac{120\degree}{M}.
\end{equation}
For $\lambda_{\mathrm{c}}/2$ spacing considered in this paper, the FDA beam pattern is periodical within $180\degree $. During a single FDA scan, the system sends $L$ OFDM symbols. The minimal number of OFDM symbols to guarantee at least one full OFDM symbol per beam is given by
\begin{equation}
    \label{eq:min_num_symb}
    L = \frac{180\degree}{\Delta \Theta} = \lceil 1.5M \rceil.
\end{equation}
The minimum duration of the angular scan corresponds to the minimum Pulse Repetition Interval (PRI) and is calculated as
\begin{equation}
    \label{eq:min_pri}
    T_{\mathrm{p}} = L T_{\mathrm{s}}  = L \frac{(K+N_{\mathrm{cp}})}{B}.
\end{equation}
The increased PRI as compared to a system without scanning reduces the maximum Doppler velocity and resolution by a factor of $L$
\begin{equation}
    \label{eq:max_doppler}
    v_{\mathrm{max}} = \frac{\lambda_{\mathrm{c}}}{4 T_{\mathrm{p}}} 
    =\frac{\lambda_{\mathrm{c}}}{4 L T_{\mathrm{s}}},
    \quad
    v_{\mathrm{res}} = \frac{\lambda_{\mathrm{c}}}{2 I T_{\mathrm{p}}} 
    = \frac{\lambda_{\mathrm{c}}}{2 I L T_{\mathrm{s}}},
\end{equation}
where $I$ is the number of aggregated scans.
The maximum unambiguous range and resolution of the FDA OFDM system are identical to that of the standard OFDM system
\begin{equation}
    \label{eq:max_range}
    r_{\mathrm{max, no CP}} = \frac{c K}{2 B},
    \quad
    r_{\mathrm{max, w CP}} = \frac{c N_{\mathrm{cp}}}{2 B},
    \quad
    r_{\mathrm{res}} = \frac{c}{2 B}.
\end{equation}
Note that, when a CP is used, the maximum unambiguous range is limited by its duration.
The minimum duration of the angular scan \eqref{eq:min_pri} determines the maximum frequency offset $\Delta f$ between antenna elements $\Delta f_{\mathrm{max}} = 1 / T_{\mathrm{p}}$. With this formulation, it becomes apparent that the antenna frequency offset is always smaller than the subcarrier spacing $\Delta f _{sc}$.

\begin{figure}[b]
    \centering
    \includegraphics[width=\columnwidth]{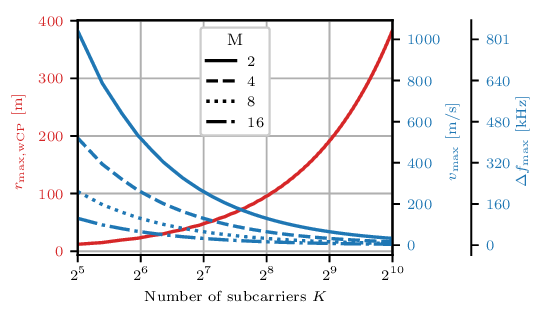}
    \caption{Scanning frequency, maximum unambiguous range and velocity as a function of a number of subcarriers $K$ for $B=100$ MHz, $f_c=60$ GHz and a selected number of antennas. ($N_{\mathrm{cp}} = 0.25K$)}
    \label{fig:fda_radar_sys_analysis}
\end{figure}

Assuming that $M$ and $B$ are fixed parameters, the radar metrics can be evaluated against the number of subcarriers $K$. Fig. \ref{fig:fda_radar_sys_analysis} shows that reducing the number of subcarriers increases the maximum scanning frequency and maximum unambiguous velocity, at the cost of reduced maximum unambiguous range. Moreover with the increasing number of antennas more OFDM symbols per scan are required, which translates to reduced $v_{\mathrm{max}}$ and $\Delta f_{\mathrm{max}}$.

\begin{figure}[t]
    \centering
    \includegraphics[width=\columnwidth]{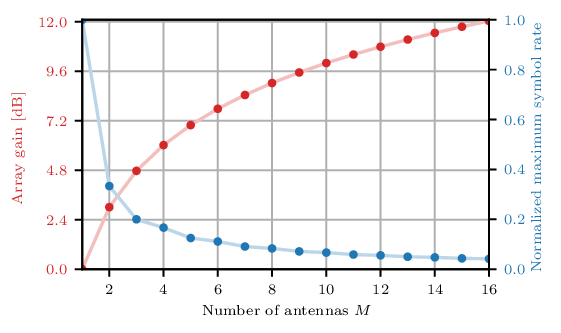}
    \caption{Antenna array gain and normalized maximum symbol rate as a function of the number of antennas. ($N_{\mathrm{cp}} = 0.25K$)}
    \label{fig:fda_comm_sys_analysis}
\end{figure}

One of the key features of the FDA is the array gain, which naturally increases with the number of antennas. The array gain allows for improved SNR at the sensing or communication receiver. As the number of antennas increases, so does the minimum number of OFDM symbols per scan \eqref{eq:min_num_symb}, due to narrowed beamwidth. As a result, a receiver located at some angle only intercepts $1/L$ of the transmitted OFDM symbols. The received symbol rate is then
\begin{equation}
    R = \frac{K B }{ L (K + N_{\mathrm{cp}})}.
\end{equation}
The achievable maximum symbol rate is inversely proportional to the number of OFDM symbols per scan. Fig.\ref{fig:fda_comm_sys_analysis} shows the array gain and maximum achievable symbol rate, normalized with respect to that of the SISO system ($L=1$).

\section{Signal model}
\label{sec:sig_proc_model}

\subsection{Modulation}
The system utilizes OFDM modulation with $K$ subcarriers with indices $k \in \{0, \ldots, K-1\}$. The complex modulation symbols $s(k)$ are chosen from a QPSK constellation. The time domain (TD) signal is obtained by performing an inverse discrete Fourier transform (IDFT) of the modulation symbols
\begin{equation}
    \label{eq:ofdm_mod}
    y(n) = \frac{1}{N} \sum_{k=0}^{N-1}s(k)e^{j2\pi\frac{n}{N}k}, 
\end{equation}
where $n \in \{0, \ldots, N-1\}$ are discrete-time indices. 
\subsection{Frequency shift}
The signal transmitted by the $m$th antenna is a scaled and frequency-shifted version of the original signal
\begin{equation}
    \label{eq:freq_shift_ofdm}
    y_m(n) = \frac{1}{\sqrt{M}} y(n) e^{j2\pi m \frac{\varepsilon}{N} n },
\end{equation}
where $\varepsilon = \frac{\Delta f}{ \Delta f_{\mathrm{sc}}} $ is the frequency offset normalized by the subcarrier spacing. The factor $1/\sqrt{M}$ normalizes the total transmit power to unity.
The frequency domain representation of the frequency-shifted signal is obtained by performing a DFT of the same length as in \eqref{eq:ofdm_mod}. Following the formulation in \cite{ofdm_ici}, the signal at antenna $m$ can be expressed as
\begin{align}
    y_m(k) &= \sum_{n=0}^{N-1} y_m(n) e^{-j2\pi \frac{k}{N} n} \nonumber \\
           &= \frac{1}{N \sqrt{M}} \sum_{l=0}^{N-1} s(l) \underbrace{\sum_{n=0}^{N-1} e^{j2\pi n \frac{(l + m\varepsilon - k)}{N}}}_{\mathrm{ICI}},
\end{align}
where $l$ is the other subcarrier index. Due to frequency shift the subcarriers are no longer orthogonal and intercarrier interference (ICI) appears. The ICI component between the $l$th and $k$th subcarrier can be written as \cite{ofdm_ici}
\begin{align}
    \label{eq:ici_coeff_closed_form}
    p_m(l-k) &= \sum_{n=0}^{N-1} e^{j2\pi n \frac{(l + m\varepsilon - k)}{N}} \nonumber \\
    &= e^{j \pi (\frac{N-1}{N})(l + m\varepsilon - k)} \frac{\sin{(\pi(l + m\varepsilon - k))}}{N \sin(\frac{\pi}{N} (l + m\varepsilon - k))}.
\end{align}
Rewriting the transmitted signal with the use of \eqref{eq:ici_coeff_closed_form} results in
\begin{align}
    y_m(k) &= \frac{1}{\sqrt{M}} \sum_{l=0}^{N-1}s(l) p_m(l-k) \nonumber \\
    &= \frac{1}{\sqrt{M}} \left( s(k) p_m(0) + \sum_{\substack{l=0 \\ l \neq k }}^{N-1} s(l) p_m(l-k) \right).
\end{align}
As a result of frequency shift, the modulation symbol at the $k$th subcarrier is scaled and rotated. Moreover, the inter-carrier interference appears as a result of the loss of orthogonality.
To include the effects of FDA beamforming on the transmitted signal, it has to be considered for some moment in time $t$ when the AF of the FDA is pointing in the direction $\theta_{\mathrm{TX}}(t)$ given by \eqref{eq:fda_angle_vs_time}. The transmitted signal for the $m$th antenna is then
\begin{align}
    y_m(k, t) = y_m(k) e^{-j 2 \pi (k \Delta f_{\mathrm{sc}} + m \Delta f) \left(\frac{m d_{\mathrm{a}}}{c} \sin{(\theta_{\mathrm{TX}}(t)}) \right)}.
\end{align}

\subsection{Communications}
The communication system is considered in the line-of-sight scenario. A single antenna receiver is located at distance $d$ and angle $\theta_{\mathrm{RX}}$. Assuming a narrow band system $\Delta f_{\mathrm{sc}} \ll f_c$, the attenuation is identical for all subcarriers and antennas. Following the per-subcarrier flat-fading assumption, the channel can be modeled as a single complex coefficient
\begin{equation}
    \label{eq:los_fd_comm_channel}
    h_m(k) = \alpha e^{-j2 \pi (k \Delta f_{\mathrm{sc}} + m \Delta f )\left( \frac{d}{c} - \frac{m d_{\mathrm{a}}}{c} \sin{(\theta_{\mathrm{RX}})} \right) }.
\end{equation}
The received signal from $m$th antenna can be written as
\begin{align}
    \label{eq:mth_ant_rx_sig}
    r_m(k, t) =& y_m(k,t) h_m(k) \nonumber \\
    =&\alpha y_m(k,t) \nonumber\\
    & e^{-j2 \pi (k \Delta f_{\mathrm{sc}} + m \Delta f ) \left( \frac{d}{c}  + \frac{m d_{\mathrm{a}}}{c} \Delta \phi(t) \right) },
\end{align}
The received signal can be further simplified by assuming that $M\Delta f \ll \Delta f_{\mathrm{sc}}$, resulting in
\begin{equation}
    \label{eq:mth_ant_rx_sig_simplified}
    r_m(k,t) = \alpha e^{j\varphi_k} y_m(k,t) e^{-j2 k \Delta f_{\mathrm{sc}} \frac{m d_{\mathrm{a}}}{c} \Delta \phi(t) },
\end{equation}
where $\varphi_k = -2 \pi k \Delta f_{\mathrm{sc}} d / c$ is the phase rotation at the $k$th subcarrier due to propagation delay and $\Delta \phi(t) = \sin{(\theta_{\mathrm{TX}}(t))} - \sin{(\theta_{\mathrm{RX}})}$. The sum of the received signals is then
\begin{equation}
    \label{eq:rx_comm_sig}
    r(k,t) =\alpha e^{j\varphi_{k}} \sum_{-(M-1)/2}^{(M-1)/2} y_m(k) e^{-j2 k \Delta f_{\mathrm{sc}} \frac{m d_{\mathrm{a}}}{c} \Delta \phi(t) },
\end{equation}

The attenuation $\alpha$ and $e^{j \varphi_{k}}$ components are subject to equalization at the receiver. The received signal formulation reveals that, when the FDA beamforming direction is not aligned with the receiver, there is an additional antenna-dependent phase rotation per subcarrier. This introduces unrecoverable signal distortion at other angles, hampering the reception quality. As the beamforming direction of the FDA is time-dependent, $\theta_{\mathrm{TX}}(t)$ the quality of the received signal suffers periodic variations.

In the simulation results section, the communication performance is quantified by the error vector magnitude (EVM) defined as
\begin{equation}
    \label{eq:evm_definition}
    \mathrm{EVM} = \frac{\sqrt{ \frac{1}{K} {\sum_{k=0}^{K-1} \lvert s(k) - r_{\mathrm{eq}}(k) \rvert }^{2} }}{|s|},
\end{equation}
where $r_{\mathrm{eq}}(k)$ is the equalized received symbol. The EVM allows to asses how the performance of the standard OFDM system is degraded by combining it with the FDA.

\subsection{Sensing}

In the sensing scenario, an isotropic scatterer $q$ is considered, located at angle $\theta_q$ and distance $d$, with complex scattering coefficient $\gamma$ and velocity $v$. The radar receiver is located at the center of the array. The observed relative velocity across all antennas is assumed to be the same. The effects of the scatterer on the received signal in the frequency domain can be formulated as 
\begin{align}
    \label{eq:radar_channel_fd}
    q_m(k, l) &= \gamma e^{-j2\pi T_{\mathrm{p}} \frac{v}{\lambda_{\mathrm{c}}}} e^{-j2\pi ( m\Delta f + k \Delta f_{\mathrm{sc}} )  \left(\frac{2d + \Delta d_m}{c} \right)}
\end{align}
The reflection signal from the $m$th antenna is
\begin{align}
    \label{eq:mth_ant_rx_sens_sig}
    x_m(k, l, t) =& y_m(k,t) q_m(k, l) \nonumber \\
    =& \frac{1}{\sqrt{M}} \gamma e^{-j2\pi l T_{\mathrm{p}} \frac{v}{\lambda_{\mathrm{c}}}} \\
    & e^{-j2\pi ( m\Delta f + k \Delta f_{\mathrm{sc}} )  \left( \frac{2d}{c}  + \frac{m d_{\mathrm{a}}}{c} \left( \sin{(\theta_{\mathrm{TX}}(t))} - \sin{(\theta_{q})} \right) \right)} \nonumber 
\end{align}
Simplifying the signal as in \eqref{eq:mth_ant_rx_sig_simplified} results in
\begin{equation}
    \label{eq:mth_ant_rx_sens_sig_simplified}
    x_m(k, l, t) = \frac{1}{\sqrt{M}} \gamma e^{j (\phi_l + 2\varphi_k)} e^{-j2\pi k \Delta f_{\mathrm{sc}} \frac{m d_{\mathrm{a}}}{c} \Delta \phi_q(t)},
\end{equation}
where $\phi_l = -j2\pi l T_{\mathrm{p}} v / \lambda_{\mathrm{c}} $ is the phase shift due to the target velocity and $\Delta \phi_q(t) = \sin{(\theta_{\mathrm{TX}}(t))} - \sin{(\theta_{q})}$.
The sum of the reflected signals is
\begin{align}
    \label{eq:rx_sens_sig}
    x(k, l, t) =& \gamma e^{j(\phi_l + 2\varphi_k)}
    \sum_{-(M-1)/2}^{(M-1)/2} y_m(k,t) e^{-j2 k \Delta f_{\mathrm{sc}} \frac{m d_{\mathrm{a}}}{c} \Delta \phi_q(t) }.
\end{align}
The channel estimate is obtained by processing the return signal with a zero-forcing estimator
\begin{equation}
    \hat{q}(k, t) = \frac{x(k,l, t)}{s(k,l)}.
\end{equation}
The range profile is obtained by performing an IDFT on the obtained channel estimate. 
By inspecting the received sensing signal in \eqref{eq:rx_sens_sig},  conclusions similar to those in the communications scenario can be drawn. An offset between the FDA beamforming angle and the target angle results in antenna-dependent subcarrier rotation, distorting the signal and increasing the sidelobe levels. Moreover, the presence of ICI due to frequency offset also contributes to the sidelobe levels and might introduce bias in target estimation.

In the simulation results section, the sensing performance is evaluated based on the integrated sidelobe level (ISL) defined as
\begin{equation}
    \label{eq:isl_definition}
    \mathrm{ISL} = \frac{ {\sum_{ k=0,\ k \neq k_{max} }^{K-1} \lvert p(k) \rvert }^{2} }{|p(k_{max})|^2},
\end{equation}
where $p(k)$ is the range profile obtained by processing the channel estimate and $k_{max}$ is the index of the maximum of the range profile.

\section{Simulation Results}

\begin{figure}[b]
    \centering
    \includegraphics[width=\columnwidth]{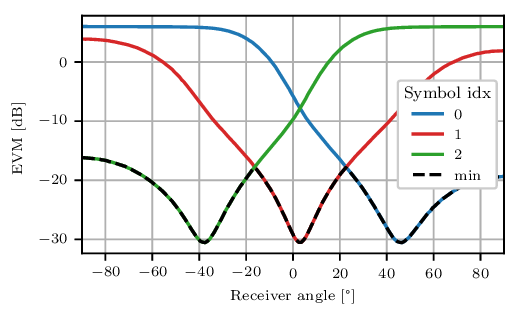}
    \caption{EVM at the communications receiver as a function of angle and OFDM symbol index, for $M=2$ antennas and $L=3$ OFDM symbols.}
    \label{fig:evm_vs_rx_ang}
\end{figure}

The performance of the FDA OFDM system was evaluated with the use of numerical simulations. The FDA is a dynamic system with time-variant beamforming and a limited number of OFDM symbols transmitted during each scan. Each OFDM symbol covers some angular extent and, as shown in Sec. \ref{sec:sig_proc_model}, the difference between the beamforming angle and target or receiver angle degrades the performance of the system. The performance dependency on target or receiver angle was investigated for a system with $M=2$ antennas and $L=3$ transmitted OFDM symbols during a scan (maximum scanning frequency).

Fig. \ref{fig:evm_vs_rx_ang} shows the EVM against the receiver angle for each of the transmitted OFDM symbols during a scan. The symbol index denotes the order in which the OFDM symbols were transmitted. 
The plot shows that the EVM greatly varies with the receiver angle. Moreover, depending on the symbol index the EVM reaches a minimum for different angles. This is a result of each OFDM symbol being beamformed in different directions as the FDA scan progresses. The system's performance across the whole angular extent is visualized by the dashed line which corresponds to the minimum of the EVM across all symbols. The per-frame angle-dependent EVM can be considered in the secure communications context. The EVM is degraded in directions other than the intended one, improving the secrecy rate and preventing an eavesdropper at other angles from intercepting the information.

\begin{figure}[b]
    \centering
    \includegraphics[width=\columnwidth]{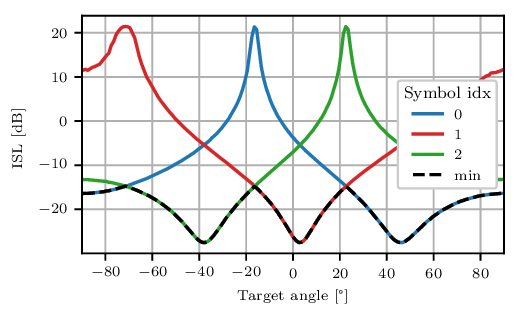}
    \caption{ISL at radar receiver as a function of target angle and OFDM symbol index, for $M=2$ antennas and $L=3$ OFDM symbols.}
    \label{fig:isl_vs_tgt_ang}
\end{figure}

Fig. \ref{fig:isl_vs_tgt_ang} presents the ISL as a function of the target angle for each OFDM symbol. Similarly, as in the communications scenario, the ISL is degraded for targets at angles other than those to which the FDA beam is directed. The system performance across all angles is visualized by the dashed line, which corresponds to the minimum of ISL across all OFDM symbols. The increased ISL for other angular directions might introduce problems as the sidelobes from closer targets at an offset angle can dominate the return of a small target at the beamforming angle.

\begin{figure}[t]
    \centering
    \includegraphics[width=\columnwidth]{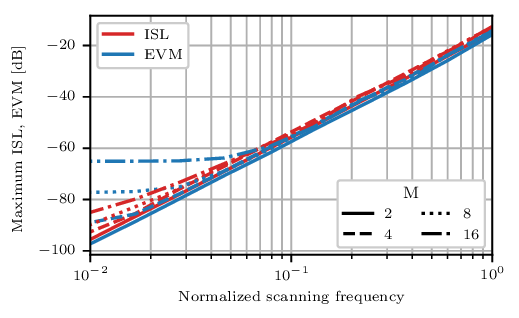}
    \caption{Maximum ISL and EVM as a function of the scanning frequency (frequency offset) between antennas - normalized to the maximum scanning frequency, for a selected number of antennas $M$.}
    \label{fig:isl_evm_vs_df}
\end{figure}

The performance of the FDA OFDM system is dependent on the angle of the target or receiver. Therefore in the following analyses, the maximum ISL and EVM values across angles are considered, which correspond to the worst-case performance. Fig. \ref{fig:isl_evm_vs_df} depicts the maximum ISL and EVM vs the scanning frequency normalized to the maximum scanning frequency for a given number of antennas. As expected, the lower the frequency offset between the antennas, the smaller the ICI contribution and both performance metrics improve. A better performance can be achieved at the cost of slower scanning. One should notice that the performance also depends on the number of antennas. As the number of antennas grows - the maximum scanning frequency is lowered - reducing the power of ICI. However, the interference from multiple antennas sums up coherently to some extent. Those two effects approximately cancel one another. Therefore, the performance of the system can be considered agnostic of the number of antennas for fast scanning. The most considerable differences in performance can be observed for a low number of antennas (i.e. $2$) or slow scanning when the performance of the system becomes limited by the number of antennas. 

\begin{figure}[t]
    \centering
    \includegraphics[width=\columnwidth]{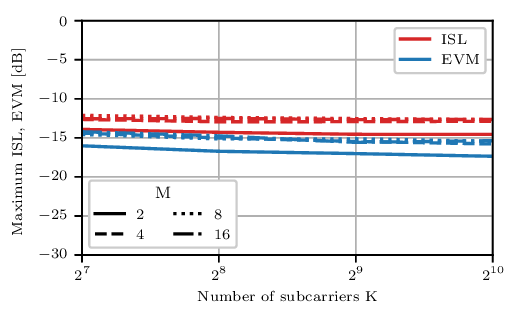}
    \caption{Maximum ISL and EVM for maximum scanning frequency as a function of the number of subcarriers for a selected number of antennas $M$.}
    \label{fig:isl_evm_vs_nsc}
\end{figure}

Fig. \ref{fig:isl_evm_vs_nsc} shows the maximum ISL and EVM as a function of the number of subcarriers. Both communications and sensing performance are unaffected by the number of subcarriers since the ratio of the frequency offset and subcarrier spacing is constant for a given $M$, regardless of the number of subcarriers.

One key property of the FDA that should not be overlooked is the ability to resolve the targets in the angular domain. From each scan, a range-angle map can be produced, which allows to resolve targets in the angular domain with a resolution similar to that of the PA \cite{phased_array_handbook}.

\section{Conclusion}
The FDAs offer an interesting alternative to manipulate the spatial properties of the signals radiated by antenna arrays. In this paper, the key features of the FDA OFDM system were discussed in the context of joint communication and sensing. The number of antennas in the FDA system effectively limits the maximum scanning frequency and throughput. On the other hand, it has a negligible impact on the system's performance and scales the array gain. By reducing the scanning frequency the ISL and EVM can be improved to meet the requirements of a specific application. For instance, a tenfold reduction scanning frequency in regard to maximum allows for a $40$ dB improvement in EVM and ISL, rendering the impact of interference due to frequency shift negligible.

\bibliographystyle{IEEEtran}
\bibliography{biblio.bib}

\begin{thebibliography}{10}
\providecommand{\url}[1]{#1}
\csname url@samestyle\endcsname
\providecommand{\newblock}{\relax}
\providecommand{\bibinfo}[2]{#2}
\providecommand{\BIBentrySTDinterwordspacing}{\spaceskip=0pt\relax}
\providecommand{\BIBentryALTinterwordstretchfactor}{4}
\providecommand{\BIBentryALTinterwordspacing}{\spaceskip=\fontdimen2\font plus
\BIBentryALTinterwordstretchfactor\fontdimen3\font minus \fontdimen4\font\relax}
\providecommand{\BIBforeignlanguage}[2]{{%
\expandafter\ifx\csname l@#1\endcsname\relax
\typeout{** WARNING: IEEEtran.bst: No hyphenation pattern has been}%
\typeout{** loaded for the language `#1'. Using the pattern for}%
\typeout{** the default language instead.}%
\else
\language=\csname l@#1\endcsname
\fi
#2}}
\providecommand{\BIBdecl}{\relax}
\BIBdecl

\bibitem{mmw_comms_it_will_work}
T.~S. Rappaport, S.~Sun, R.~Mayzus, H.~Zhao, Y.~Azar, K.~Wang, G.~N. Wong, J.~K. Schulz, M.~Samimi, and F.~Gutierrez, ``Millimeter wave mobile communications for {5G} cellular: It will work!'' \emph{IEEE Access}, vol.~1, pp. 335--349, 2013.

\bibitem{fda_tma_overview}
W.-Q. Wang, H.~C. So, and A.~Farina, ``An overview on time/frequency modulated array processing,'' \emph{IEEE Journal of Selected Topics in Signal Processing}, vol.~11, no.~2, pp. 228--246, 2017.

\bibitem{fda_mimo_techniques}
P.~F. Sammartino, C.~J. Baker, and H.~D. Griffiths, ``Frequency diverse {MIMO} techniques for radar,'' \emph{IEEE Transactions on Aerospace and Electronic Systems}, vol.~49, no.~1, pp. 201--222, 2013.

\bibitem{fda_beampattern_analysis}
Z.~Ahmad, M.~Chen, and S.-D. Bao, ``Beampattern analysis of frequency diverse array radar: a review,'' \emph{EURASIP Journal on Wireless Communications and Networking}, vol. 2021, no.~1, p. 189, Nov 2021.

\bibitem{fda_lfm_experimentation}
P.~M. McCormick, A.~Jones, N.~Kellerman, B.~Mathieu, and A.~Mertz, ``Experimental demonstration of a low-complexity multiple-input single-output frequency diverse array framework,'' in \emph{2023 IEEE Radar Conference (RadarConf23)}, 2023, pp. 1--6.

\bibitem{fda_ofdm_sc_per_antenna}
Y.~Ding, J.~Zhang, and V.~Fusco, ``Frequency diverse array {OFDM} transmitter for secure wireless communication,'' \emph{Electronics Letters}, vol.~51, no.~17, pp. 1374--1376, 2015.

\bibitem{fda_ofdm_jcns_chirped}
H.~Huang and W.-Q. Wang, ``{FDA-OFDM} for integrated navigation, sensing, and communication systems,'' \emph{IEEE Aerospace and Electronic Systems Magazine}, vol.~33, no. 5-6, pp. 34--42, 2018.

\bibitem{fma_tx_jcns}
N.~S. Mannem, E.~Erfani, T.-Y. Huang, and H.~Wang, ``A mm-wave frequency modulated transmitter array for superior resolution in angular localization supporting low-latency joint communication and sensing,'' \emph{IEEE Journal of Solid-State Circuits}, vol.~58, no.~6, pp. 1572--1585, 2023.

\bibitem{fda_jcns_index_modulation}
M.~Li and W.-Q. Wang, ``Joint radar-communication system design based on {FDA-MIMO} via frequency index modulation,'' \emph{IEEE Access}, vol.~11, pp. 67\,722--67\,736, 2023.

\bibitem{phased_array_handbook}
R.~Mailloux, \emph{Phased Array Antenna Handbook, Third Edition}.\hskip 1em plus 0.5em minus 0.4em\relax Artech House Publishers, 2017.

\bibitem{balanis_antenna_book}
C.~Balanis, \emph{Antenna Theory: Analysis and Design}, ser. Wiley-interscience.\hskip 1em plus 0.5em minus 0.4em\relax Wiley Interscience, 2005.

\bibitem{ofdm_ici}
Y.~Zhao and S.-G. Haggman, ``Intercarrier interference self-cancellation scheme for {OFDM} mobile communication systems,'' \emph{IEEE Transactions on Communications}, vol.~49, no.~7, pp. 1185--1191, 2001.

\end{thebibliography}

\end{document}